\documentclass[pre,reprint,floatfix]{revtex4-2}  % Initially 9 pt.
\usepackage{graphicx}
\usepackage[T1]{fontenc}
\usepackage[utf8]{inputenc}
\usepackage{xcolor}

\newcommand{\rem}[1]{\textcolor{red!50!gray}{}}

\definecolor{cream}{RGB}{222,217,201}

\begin{document}

\title{Tuning local microstructure of colloidal gels by ultrasound-activated deformable inclusions} 
\date{\today}

\author{Brice Saint-Michel}
\altaffiliation{Present address : Univ Gustave Eiffel, CNRS, Ecole des Ponts ParisTech, UMR 8205 Laboratoire Navier, 5 Boulevard Descartes, CEDEX 2, F-77454 Marne-la-Vall{\'e}e, France}
\affiliation{Department of Chemical Engineering, Delft University of Technology, Delft 2629 HZ, the Netherlands}

\author{George Petekidis}
\affiliation{IESL – FORTH and Department of Material Science and Technology, University of Crete, GR – 71110, Heraklion, Greece}

\author{Valeria Garbin}
\email{v.garbin@tudelft.nl}
\affiliation{Department of Chemical Engineering, Delft University of Technology, Delft 2629 HZ, the Netherlands}

\begin{abstract}
Colloidal gels possess a memory of previous shear events, both steady and oscillatory. This memory, embedded in the microstructure, affects the mechanical response of the gel, and therefore enables precise tuning of the material properties under careful preparation. Here we demonstrate how the dynamics of a deformable inclusion, namely a bubble, can be used to locally tune the microstructure of a colloidal gel. We examine two different phenomena of bubble dynamics that apply a local strain to the surrounding material: dissolution due to mass transfer, with a characteristic strain rate of $\sim10^{-3}$~s$^{-1}$; and volumetric oscillations driven by ultrasound, with a characteristic frequency of $\sim10^4$~s$^{-1}$. 
% --
We characterise experimentally the microstructure of a model colloidal gel around bubbles in a Hele-Shaw geometry using confocal microscopy and particle tracking. In bubble dissolution experiments, we observe the formation of a pocket of solvent next to the bubble surface, but marginal changes to the microstructure. In experiments with ultrasound-induced bubble oscillations, we observe a striking rearrangement of the gel particles into a microstructure with increased local ordering. High-speed bright-field microscopy reveals the occurrence of both high-frequency bubble oscillations and steady microstreaming flow; both are expected to contribute to the emergence of the local order in the microstructure. These observations open the way to local tuning of colloidal gels based on deformable inclusions controlled by external pressure fields.
\end{abstract}

\maketitle

\section{Introduction}

Colloidal gels have been intensely studied in the last twenty years owing to the richness of their phase transitions and structure at equilibrium and out of equilibrium, because they are easy to model in numerical simulations, and also because particles are readily observed and their interactions are easily tuned in experiments~\cite{Zaccarelli2007,Petekidis2021}. This rich, yet generic behaviour guides our understanding of more complex formulations used in industrial applications, such as carbon black gels~\cite{Grenard2014}, fumed silica~\cite{Walls2003}, natural rubber latex gels~\cite{deOliveiraReis2019} and cementious materials~\cite{Ioannidou2016}.

Depending on the particle interaction strength compared to the thermal energy $k_{\rm B} T$, gels of attractive colloidal particles may phase separate~\cite{Lu2008,Whitaker2019}, dynamically arrest~\cite{Weitz1984}, or follow more complex pathways~\cite{Rouwhorst2020} towards forming space-spanning networks. The structure of such networks, which coarsens with time~\cite{Zia2014}, dictates whether and how the network may support shear stresses in the linear regime~\cite{Bouzid2018,Whitaker2019} and governs their rheological behaviour for larger deformation~\cite{Hsiao2012,Koumakis2015,Ruiz-Franco2020}. Oscillatory shear induces unique structures in complex fluids~\cite{Corte2008} and in particular in colloidal gels~\cite{Smith2007,Moghimi2017,Schwen2020}, providing additional control over their microstructure and rheological properties~\cite{Moghimi2017}. For instance, shear-induced crystallisation also imparts remarkable acoustic~\cite{Cummer2016} or photonic properties for advanced material applications~\cite{Soukoulis2011}. 

Classical experimental devices impose oscillatory shear at a relatively low frequency, both in the linear~\cite{Vermant1997,Moghimi2019} or in the non-linear regimes~\cite{Smith2007,Moghimi2017}. Interestingly, more recent experiments have evidenced that applying small oscillations at high-frequency can have a strong impact on the rheology of attractive colloidal gels~\cite{Gibaud2020} and shear-thickening suspensions~\cite{Sehgal2019}, which can be obtained for instance by propagation of ultrasound waves in the medium. Ultra-small angle X-ray scattering measurements~\cite{Gibaud2020} show direct evidence that fractal aggregates forming the gels break following the application of high-power ultrasound, leading to an increased fluidity. Materials possessing a microstructure sensitive to the application of transverse ultrasound oscillations are called \emph{rheoacoustic}, and mainly encompass fragile materials with a limited elastic plateau.

We propose to extend the applicability of ultrasound manipulation of complex fluids by embedding deformable inclusions that can be activated by ultrasound, namely gas bubbles. Gas bubbles can deform either by mass transfer (passive dissolution) or can be actively driven into volumetric oscillations under ultrasound excitation \cite{SaintMichel2020}. These oscillations have already been shown to drive dynamic self-assembly in colloid monolayers on the surface of  bubbles~\cite{Huerre2018}, and to destroy such monolayers for large-amplitude forcing~\cite{Poulichet2015b}. The motion of the spherical bubble surface also applies a local extensional strain onto the surrounding medium. Since the resonance of the bubble is affected by the local mechanical properties of the surrounding media, bubble dynamics are also used to locally probe the rheology of soft materials~\cite{Jamburidze2017, Estrada2018, SaintMichel2020b}. For rheoacoustic materials, the strain applied by the bubbles may also exceed the local yield strain and induce irreversible changes to their microstructure. Since this strain quickly decays away from the bubble, this mechanism could be very well-suited to locally tune and disrupt the microstructure of soft materials. 

In this paper we explore the evolution of the microstructure of a depletion colloidal gel around a bubble under passive dissolution and under ultrasound-driven bubble oscillations. Using confocal microscopy, we show that bubble dissolution usually leaves a large pocket of solvent free of particles at the initial site of the bubble. In contrast, bubble oscillations induce a local, preferential orientation of the bonds in the gel surrounding the bubble, which we quantify by introducing a suitable order parameter. High-speed, bright-field microscopy data provide some further insights into the dynamics of deformation of the gel during bubble oscillations, and point to a possible mechanism at the origin of the observed local ordering.

\section{Materials and Methods}

\begin{figure}[h]
    \centering
    \includegraphics[width=7cm]{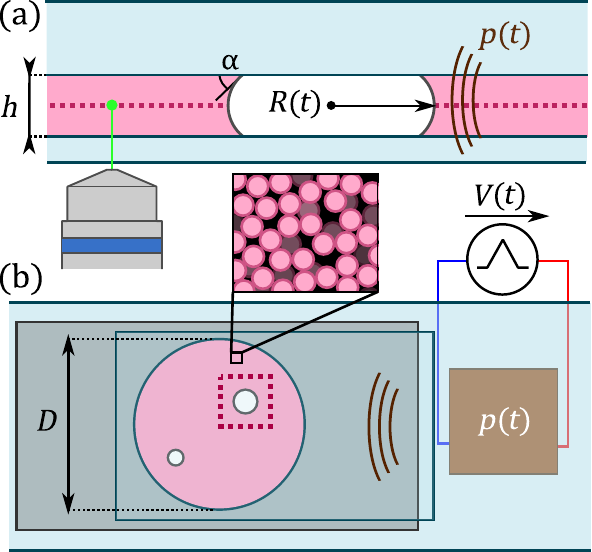}
    \caption{Schematic of the experimental setup (not to scale) shown (a) in side view and (b) in bottom view. The fluorescent colloidal gel (pink) lies between a glass slide and a glass coverslip in a cylindrical pit of height $h = 95~\mu$m and diameter $D = 9$~mm. Bubbles are squeezed between the two plates and driven into oscillations by the acoustic wave, $p(t)$, generated by a piezoelectric transducer (brown element). The gel is imaged in  the equatorial plane of the bubble using a confocal microscope.}
    \label{fig:SCHEMATIC}
\end{figure}

\paragraph*{Fluorescent colloidal gel}

The fluorescent colloidal gel consists of a colloid-polymer mixture similar to that used by Moghimi {et al.}~\cite{Moghimi2017}. It consists of PMMA particles of nominal radius $541$~nm, doped with fluorescent dye (NBD), sterically stabilised by short grafted chains ($\simeq 10$~nm) of poly-hydroxystearic acid chains. The particles are suspended in squalene (optical index $n = 1.494$, density $\rho_{\rm l} = 858$~kg\,m$^{-3}$, surface tension $\gamma = 0.029$~N\,m$^{-1}$, dynamic viscosity $\eta = 0.012$~Pa\,s) to achieve refractive index matching. Linear polybutadiene (molecular weight $M_w = 1.243 \, 10^6$~g.mol$^{-1}$, concentration $c = 5.22\,10^{-2}$~g\,L$^{-1}$), dispersed into the squalene, induces depletion interactions between the particles~\cite{Asakura1954}. The range of the attractive interaction is $\xi = 22.4$~nm, equal to the gyration radius of the linear chains, and its magnitude at contact, $U_{\rm dep} = -20 k_{\rm B}T$, is strong. The volume fraction of the particles is $\varphi=44\%$. The rheological properties of the gel are shown in SI Figure 1 : the amplitude sweep conducted at $1$~Hz shows that the gel is a weak solid of modulus $G' \simeq 10$~Pa and exhibits two-step yielding~\cite{Koumakis2011}. The first yield point, where particle rearranging starts, occurs at a yield strain around $\gamma_{\rm Y} = 1.6\%$ corresponding to a yield stress $\sigma_{\rm Y} \simeq 0.2$~Pa.

\paragraph*{Sample preparation}

The sample geometry is shown in Figure~\ref{fig:SCHEMATIC}. It consists of a large microscope slide ($76 \times 52$~mm$^2$), onto which we stick a spacer with a circular hole of diameter $9$~mm. We then obtain a pit of height $h = 95~\mu$m and volume $6~\mu$l. Samples are prepared by placing a small gel drop in the pit using a pipette. The pipette is then used to incorporate small bubbles into the gel. The sample is closed using a large glass coverslip ($24 \times 50$~mm$^2$) and sealed immediately using cyanoacrylate glue. Bubbles may assume either a spherical cap shape or a pancake shape depending on their volume and the contact angle $\alpha$ [see Fig.~\ref{fig:SCHEMATIC}(a)] between the solvent, the glass and air. Microscopic observations begin as soon as the sample is sealed and connected to the acoustic excitation setup described below. We only consider bubbles that are sufficiently isolated, i.e., at least two diameters away from other bubbles and the sample edge. 

\paragraph*{Ultrasound excitation}
A piezoelectric transducer (P-141.10, Physik Instrumente) is glued onto the other side of the glass slide to apply the ultrasound excitation [see Figure~\ref{fig:acoustics_observations}(a)]. We drive this transducer using a waveform generator (33210A, Agilent) and a linear power amplifier (AG 1021, T\&C Power Conversion) allowing us to choose the excitation amplitude through the applied voltage $V$, its frequency $15 \leq f \leq 21$~kHz and its duration, given by the number of applied cycles $N=10^2-10^6$. Due to the very small size of our geometry, we cannot directly measure the pressure field applied to the bubble and the gel. We have checked in control experiments (shown in SI Figure 2) that the pressure field itself has no impact on the gel in the absence of bubbles.

\paragraph*{Real-time and high-speed brightfield imaging setups}
\label{sec:brightfield}

We image the samples in real-time brightfield microscopy using a Zeiss inverted microscope, working (unless otherwise noted) with a 40$\times$ Fluar oil-immersion objective and a Axiocam MRm camera (Carl Zeiss, $1388 \times 1040$ pixels). We use this mode to focus on the equatorial plane of relatively large bubbles assuming a flattened shape (see Figure~\ref{fig:SCHEMATIC}(a)), before switching to confocal acquisitions (see below)). The pixel size in brightfield mode is equal to $0.161~\mu$m, and images are acquired at 4.5 frames per second, with the exposure time equal to the time interval between pictures; this frame rate is naturally too small to resolve the bubble oscillations.

We use a second brightfield setup for high-speed video microscopy of bubble oscillation dynamics. It consists of an inverted microscope (IX71, Olympus) coupled to a 40$\times$ objective and a high-speed camera (Fastcam SA5, Photron). The resolution in this configuration is $0.5~\mu$m, insufficient to resolve the individual gel particles. We acquire high-speed recordings at a rate, e.g. $7000$ frames per second for $f=20.7$~kHz, that is insufficient to resolve one bubble oscillation cycle, but work with a very short exposure time of $1/150\,000~$s to suppress motion blur. We then obtain time-aliased acquisitions allowing us to characterise both the oscillation and long-term dynamics of the bubble and the surrounding gel provided that the bubble oscillation amplitude does not vary with time (see SI Section 7 for more details). 

\paragraph*{Confocal Imaging}
\label{sec:confocal}
Confocal images are taken with the real-time microscope described in Section~\ref{sec:brightfield} coupled to a Zeiss LSM 710 confocal unit. We choose a region of interest of $3072 \times 3072$ pixels with a resolution of $0.1~\mu$m per pixel. This large region of interest, amounting to $\sim 300 \times 300~\mu$m$^2$, is needed to image the whole bubble cross-section and its surroundings and resolve the particles at the same time. The time for acquisition of a slice is between $8$ and $10$ seconds, insufficient to resolve the bubble oscillation dynamics during the application of ultrasound. Acquiring $z$-resolved stacks of images to examine the three-dimensional structure of the gel is also difficult due to bubble dissolution (see Section~\ref{sec:dissolution}). We then examine the microstructure of the gel in the equatorial plane of the bubble (the plane of interest) either in a time-resolved manner for passive bubble dissolution, or in a before/after fashion for ultrasound experiments. The stacks are analysed using standard tracking routines~\cite{Allan2021} to detect particles and compute their centres~\cite{Crocker1996}. The algorithm correctly recovers almost all particles in the field of view, as shown for a sample picture in SI Figure 3.

\paragraph*{Kinematics of pancake bubbles and surrounding medium}

Bubbles that are flattened like a pancake between the two glass slides\cite{MekkiBerrada2016} usually show a bright central region surrounded by a dark annular region in brightfield microscopy [Figure~\ref{fig:microstreaming}(c-e)] and simply appear as a black circle in confocal microscopy [Figure~\ref{fig:dissolution_observations}(b)]. We define the projected radius $R(t)$ as the distance between the bubble centre and the outer edge of the annular or circular black region. In addition, knowing the thickness of the dark ring in brightfield microscopy and the height of the sample provides an estimate of the contact angle between the solvent and glass, $\alpha$, found to be close to $0^\circ$. During ultrasound excitation, the projected bubble radius evolves as $R(t)=R_0 [ 1 + x \cos (2 \pi f t) ]$ with $x$  the non-dimensional amplitude of oscillations.

Pancake bubbles, like spherical bubbles, impose a velocity field to their surroundings when their radius $R(t)$ changes. This field is prescribed by the no-penetration condition at the bubble surface and the incompressibility of the surrounding medium, leading to ${\bf v} = \left (\dot{R} R^2/r^2  \right ) {\bf e_r}$ for spherical bubbles. This velocity field produces a purely extensional strain field decaying like $r^{-3}$ away from the bubble~\cite{SaintMichel2020}. Bubbles then completely prescribe how the constituents of the surrounding medium are strained, as long as it behaves as a homogeneous medium. Due to their geometry, pancake bubbles corresponds more closely to the ideal case of cylindrical bubbles~\cite{MekkiBerrada2016} for which ${\bf v} = \left (\dot{R}R/r \right ) {\bf e_r}$ and the strain field decays as $r^{-2}$. We expect this relation to hold close to the bubble equator, where we perform our acquisitions -- unless explicitly stated otherwise.

\section{Dimensional analysis of bubble and gel dynamics}

Bubbles in complex fluids generally assume complex shapes due to the interplay between buoyancy, surface tension and static (yield) or dynamic internal (visco-elastic) stresses in the surrounding material even at low Reynolds numbers. This interplay means that the gel structure may be disrupted due to bubble dissolution or oscillations. In this section, we use dimensional analysis to evaluate the physical processes governing the bubble shape and its dynamics, and those inducing changes in the gel structure.

\subsection{Dimensionless groups governing bubble shape, dissolution and rise}
\label{sec:dimensionless_bubble}

The shape of the bubbles are unaffected by gravity, as can be shown comparing the relative magnitude of surface tension and gravity effects using the Bond-E\"otv\"os number Bo:
\begin{equation}
    {\rm Bo} = \frac{\rho_{\rm l} g h^2}{\gamma} = 3 \times 10^{-3}\,,
\end{equation}
in which $\rho_{\rm l}$ is the density of squalene, $\gamma$ is the surface tension between squalene and air $h$ is the height of the sample geometry and $g$ is the acceleration of gravity.

Internal stresses in the colloidal gel might also lead to non-spherical bubble shapes even at low Bond-E\"otv\"os numbers. We then need to examine the elasto-capillary number El of the bubbles in the gel, defined as the ratio between surface energy and linear elastic energy of the gel. Knowing that the contact angle $\alpha$ between glass and squalene is close to $0^\circ$, we assume that the typical radius of curvature of the bubbles is close to $h/2$ when they are squeezed between the two plates, i.e. when they assume a pancake shape. In this case, we obtain:
\begin{equation}
    {\rm El} = \frac{G h}{2 \gamma} = 0.02\,,
\end{equation}
$G$ being the elastic modulus of the gel, defined as the low frequency plateau of $G'$. The small value of El confirms that surface tension solely governs the bubble shape, which will then assume either a symmetrical pancake shape or a spherical cap shape when their radius $R$ becomes small enough that they are no longer in contact with both walls.

Dimensionless groups based on the yield stress of the colloidal gel, $\sigma_{\rm Y}$, provide information on whether the surrounding medium behaves predominantly as an elastic solid or a viscous fluid. The plasto-capillary number Pl examines whether the yield stress is large enough to arrest bubble dissolution~\cite{SaintMichel2020}. The value for our system is:
\begin{equation}
    {\rm Pl} = \frac{\sigma_{\rm Y} h}{2 \sqrt{3} \gamma} = 2 \times 10^{-4},
\end{equation}
hence we expect that the yield stress of the gel will not affect the bubble dissolution dynamics at all. We also compute the yielding number $Y^{-1}$ to examine whether the yield stress is sufficient to prevent bubble rise:
\begin{equation}
    \label{eq:yieldingnumber}
    Y^{-1} = \frac{2 \sqrt{3} {\rm Bo}}{{\rm Pl}} =  \frac{\rho_{\rm l} g h}{\sigma_{\rm Y}} = 4.66\,,
\end{equation}
where we have considered a cylindrical bubble of height $h$. The value of the yielding number is below the critical value $Y^{-1}_{\rm c} = 5.1$ at which bubbles spontaneously rise in the fluid according to numerical simulations~\cite{Dimakopoulos2013}. Small, unconfined bubbles should therefore not rise in the gel.

\subsection{Dimensionless groups affecting colloidal gel structure}
\label{sec:dimensionless_gel}

To examine the effect of bubble dissolution and oscillations on the gel structure, we can compare the diffusivity of a single particle in the pure solvent with its typical motion due to the bubble oscillations, corresponding to the Péclet number of our experiments. For a typical value for the relative bubble oscillation amplitude $x = 0.01$, we have:
\begin{equation}
    {\rm Pe}^{\rm osc} = \dot\epsilon_{\rm max} \tau_{\rm B} = \frac{2 \pi^2 f \eta a^3 x}{k_{\rm B} T} = 1.3 \times 10^3\,,
\end{equation}
where $\tau_{\rm B} = \pi a^3 \eta / k_{\rm B} T$ is the time over which a free particle diffuses over its radius due to Brownian motion, $\dot\epsilon_{\rm max}$ is the (maximum) strain rate applied onto the fluid close to the bubble, $\eta = 0.012$~Pa\,s is the solvent viscosity, $a \sim 0.5~\mu$m is the particle radius, $15~\text{kHz} \leq f \leq 21~\text{kHz}$ is the ultrasound excitation frequency, $T = 293$~K is the temperature, and $k_{\rm B}$ is the Boltzmann constant. Free particles in the gel will then mostly explore their environment through the strain field induced by bubble oscillations rather than through Brownian motion.

We also compute the Péclet depletion number (also called Mason number), which compares the viscous drag force of the gel particles to the interparticle attractive force originating from depletion interactions~\cite{Koumakis2011,Varga2019,Wagner2021}:
\begin{equation}
    {\rm Pe}_{\rm dep}^{\rm osc} = \frac{24 \pi^2 \eta f a^2 \xi x}{|U_{\rm dep}|} = 12 {\rm Pe^{\rm osc}} \frac{k_{\rm B} T}{|U_{\rm dep}|} \frac{\xi}{a} \simeq 40,
\end{equation}
considering the particle interaction magnitude at contact $U_{\rm dep} = - 20 k_{\rm B} T$ and a range of interaction $\xi = 22.4$~nm. These values indicate that the the strain applied by bubble oscillations can alter the microstructure of the gel, and in particular that bonds between pairs of particles may be broken by the oscillatory viscous stresses generated by the ultrasound excitation.

\clearpage
\section{Results and Discussion}

\subsection{Effect of bubble dissolution on gel structure}
\label{sec:dissolution}

\begin{figure*}
    \centering
    \includegraphics[width=17cm]{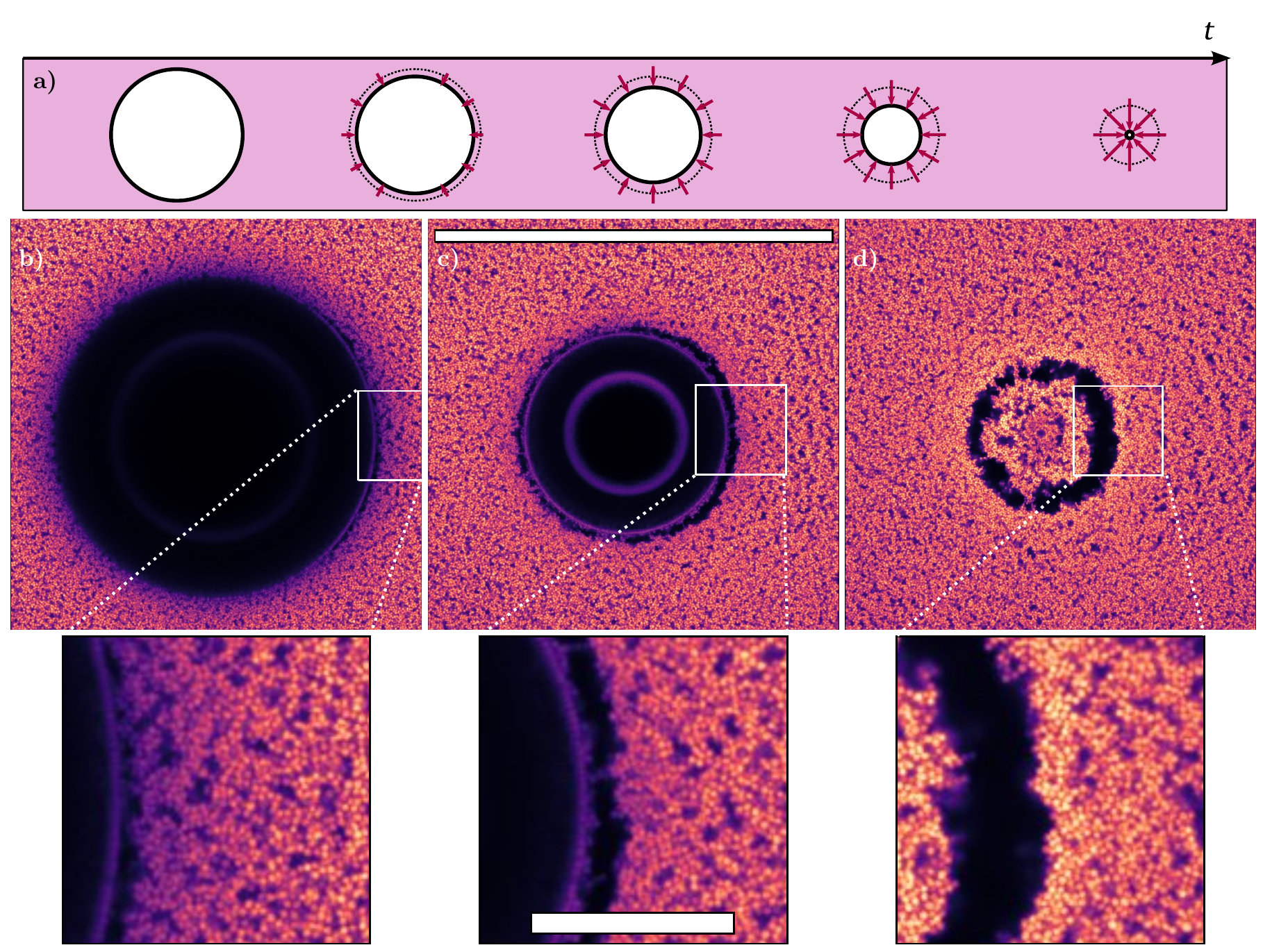}
    \caption{Effect of bubble dissolution on the colloidal gel. (a) Schematic representation of the dissolution kinematics: the bubble shrinks increasingly fast, straining the gel close to the bubble surface. (b-d) Horizontal confocal slices taken at the initial equatorial plane of the bubble at times $t=0$, $t=600$~s and $t=1020$~s. The bubble, visible in Panels (b-c), has dissolved in Panel (d), leaving a pocket of solvent and a shell of particles. The scale bar for the main panels is $200~\mu$m. Insets correspond to zoomed-in views of the gel microstructure close to the bubble (or the solvent pocket) edge. The scale bar for the insets is $30~\mu$m. See SI Video 1.}
    \label{fig:dissolution_observations}
\end{figure*}

Bubbles dissolve in the gel samples (in the absence of any acoustic excitation) over a time scale of 10-20 minutes, as schematically shown in Figure~\ref{fig:dissolution_observations}(a). During the initial stages of bubble dissolution, the colloidal particles remain very close to the interface of the bubble as seen in the inset of Figure~\ref{fig:dissolution_observations}(b). An annular pocket of pure solvent progressively grows as the bubble edge recedes [see Figure~\ref{fig:dissolution_observations}(c) and inset], until the bubble has completely dissolved and leaves behind only the solvent pocket [Figure~\ref{fig:dissolution_observations}(d)]. A layer of colloidal particles is sometimes attached to the surface of the bubble [see insets of Figure~\ref{fig:dissolution_observations} (b-c)]; these particles eventually detach from both the bulk gel and the bubble and show up as rafts within the solvent pocket after complete bubble dissolution, as shown in Figure~\ref{fig:dissolution_observations}(d). SI Video 1 shows additional steps of the dissolution of this particular bubble.  

\begin{figure*}
    \centering
    \includegraphics[width=17cm]{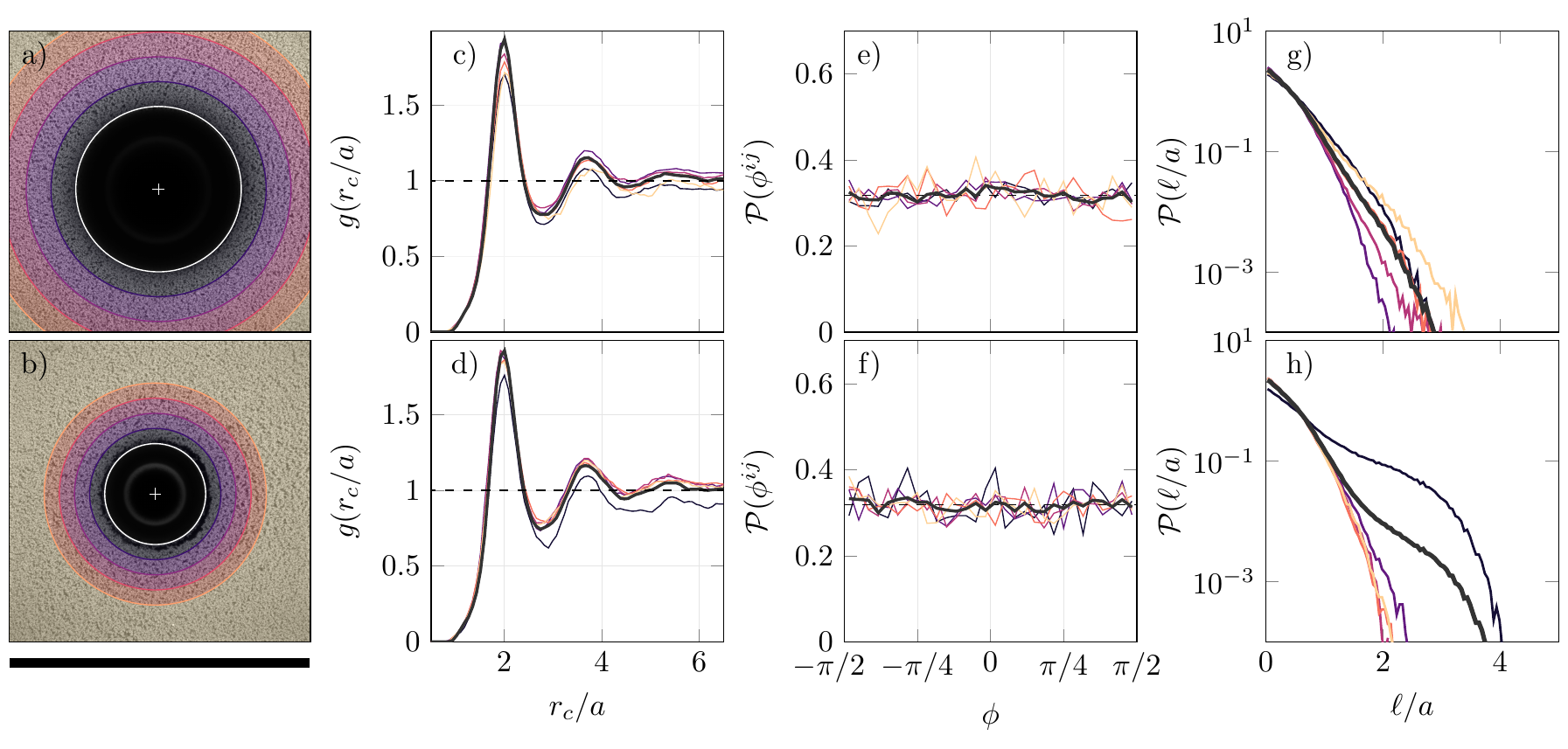}
    \caption{Evolution of the gel microstructure during bubble dissolution. Panels (a-b) correspond to Panels (b-c) of Figure~\ref{fig:dissolution_observations} and overlays indicate sectors in $r$ over which structural quantities are averaged in other panels. The scale bar is $300~\mu$m.  Panels (c-d) show the global particle pair correlation function (thick, dark grey) while sector-averaged pair correlation functions are shown as thinner, coloured lines. The inter-particle distance $r_c$ is normalised by $a$, half of the $r_c$ value where the first peak of $g(r_c)$ is observed (constant throughout this acquisition). Panels (e-f) correspond to the bond orientations $\phi^{ij}$, using the same colour conventions as in Panels (c-d). Panels (g-h) show the distribution of void sizes $\ell/a$ in the gel (see Figure~\ref{fig:A_DEFINITIONS} for details) using the same colour conventions as Panels (c-f).}
    \label{fig:dissolution_structure}
\end{figure*}

Figure~\ref{fig:dissolution_structure} shows the evolution of several indicators of gel microstructure during dissolution, defined in detail in Appendix~\ref{appendix:structure}. These quantities are either averaged over the full image, or divided into annular sectors of width $0.3 R$ starting from $R$ to $2.5 R$, as indicated in Figure~\ref{fig:dissolution_structure}(a-b). As these quantities are computed on a single confocal slice near the bubble equator, they are two-dimensional by nature and cannot capture three-dimensional structuring effects. Figure~\ref{fig:dissolution_structure}(c-d) shows the classical pair correlation function $g(r_c/a)$. We observe a clear peak around $r_c = 2a$, that we use to define the experimental size of the particles $a = 0.66~\mu$m. The pair correlation function also shows secondary peaks, but no long-range order is observed beyond $r_c > 6a$. Comparison of Figures~\ref{fig:dissolution_structure}(c-d) shows that $g(r_c/a)$ is unaffected by bubble dissolution, except very close to the bubble at the later stages, where the buildup of the solvent pocket leads to a locally lower particle density. Figure~\ref{fig:dissolution_structure}(e-f) shows the distribution of the bond orientations $\phi^{ij}$, as defined in Figure~\ref{fig:A_DEFINITIONS} (b). Both panels show that the bonds do not assume a preferred orientation relative to the surface of the bubble at any stage of bubble dissolution, despite the relatively large extensional strain locally applied onto the gel during this process. Figure~\ref{fig:dissolution_structure}(g-h) shows the distribution of the void sizes $\ell$ in the gel defined in Figure~\ref{fig:A_DEFINITIONS} (c). The presence of the solvent pocket, visible in Figure~\ref{fig:dissolution_structure}(b), leads to a very wide distribution of $\ell / a$ close to the bubble, while the voids are the smallest in the sector adjacent to that corresponding to the solvent pocket. The local particle fraction during dissolution is shown in SI Figure~4: the data captures the growth of the solvent pocket with time, appearing here as a local dip in volume fraction; it also shows a slight, local increase of the particle volume fraction right outside the solvent pocket at later times.

Bubble dissolution can in principle be arrested by the bulk yield stress of the surrounding material~\cite{Kloek2001,SaintMichel2020}, by the interfacial yield stress of particles adsorbed on the bubble surface~\cite{Beltramo2017}, or both~\cite{Saha2020}. In our system, the yield stress of the colloidal gel is too low to arrest bubble dissolution, as predicted in Section~\ref{sec:dimensionless_bubble}, and the interfacial particle layer cannot prevent dissolution either, similarly to what has been reported in Newtonian solvents under-saturated with gas~\cite{Poulichet2015,Achakulwisut2017}.

\subsection{Effect of bubble oscillations on the gel structure}

\begin{figure*}
    \centering
    \includegraphics[width=17cm]{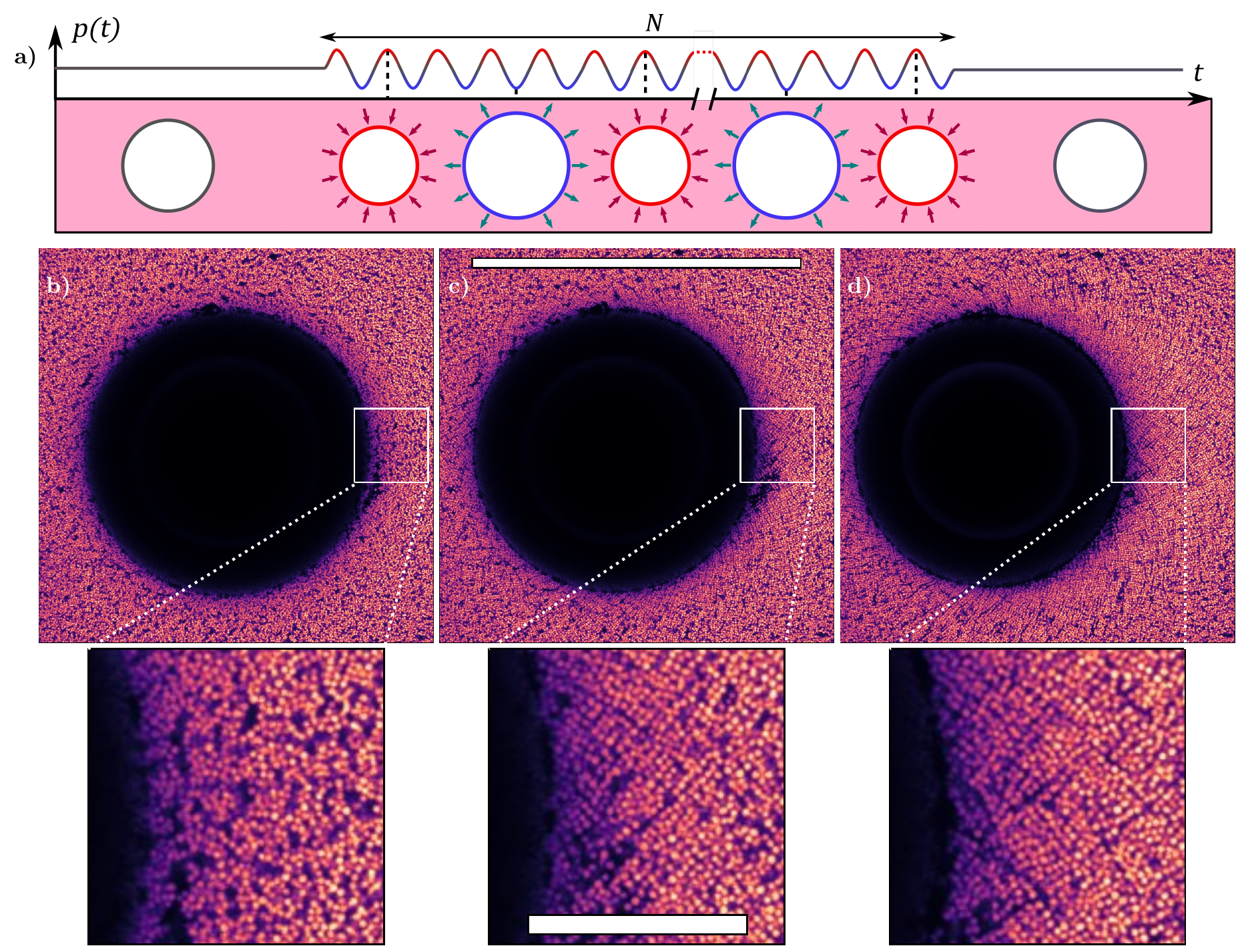}
    \caption{Acoustic excitation of a bubble. Panel (a): schematic representation of the fast oscillations into which the bubble is driven following the application of an ultrasound excitation. The number of oscillations is in the range $10^2 \leq N \leq 10^6$. The surrounding gel is periodically strained by the bubble oscillations. Panels (b-d): confocal acquisitions performed just before [Panel (b)], just after a single round [Panel (c)] and just after multiple rounds [Panel (d)] of ultrasound excitation. The scale bar for the main panels is $200~\mu$m. Insets correspond to zoomed-in views of the gel microstructure close to the bubble edge. The scale bar for the insets is $30~\mu$m.}
    \label{fig:acoustics_observations}
\end{figure*}

Figure~\ref{fig:acoustics_observations}(a) recalls the oscillatory and local nature of the strain field following the application of ultrasound, as described in Section~\ref{sec:brightfield}. We show in Figure~\ref{fig:acoustics_observations}(b) a confocal image in the equatorial plane of the bubble, just before the application of ultrasound. Figure~\ref{fig:acoustics_observations}(c) shows the same bubble just after the application of a strong acoustic pulse ($V = 180$~V peak-to-peak, $N=500$). A markedly different microstructure is visible around the bubble. Upon closer inspection (see insets of Figure~\ref{fig:acoustics_observations}), we observe that the structure of the gel close to the bubble edge has become ordered, with well-aligned particles along two preferred directions, an effect that is not observed during bubble dissolution (see Figure~\ref{fig:dissolution_observations}). A confocal $z$-stack, shown in Appendix~\ref{appendix:structure} [Figure~\ref{fig:A_ZSTACK}], indicates that the ordering appears maximal in the horizontal planes close to the bubble equator. Applying two additional pulses (again with $V = 180$~V$_{\rm pp}$, $N=500$), seems to further increase particle ordering [see Figure~\ref{fig:acoustics_observations}(d) and inset], as will be confirmed by the microstructural analysis that follows. It should be noted that the slight displacement of the centre of the bubble seen in Figures~\ref{fig:acoustics_observations}(c) and (d), which is due to acoustic radiation forces, does not have a measurable effect on the rearrangement of the microstructure. 

\begin{figure*}
    \centering
    \includegraphics[width=17cm]{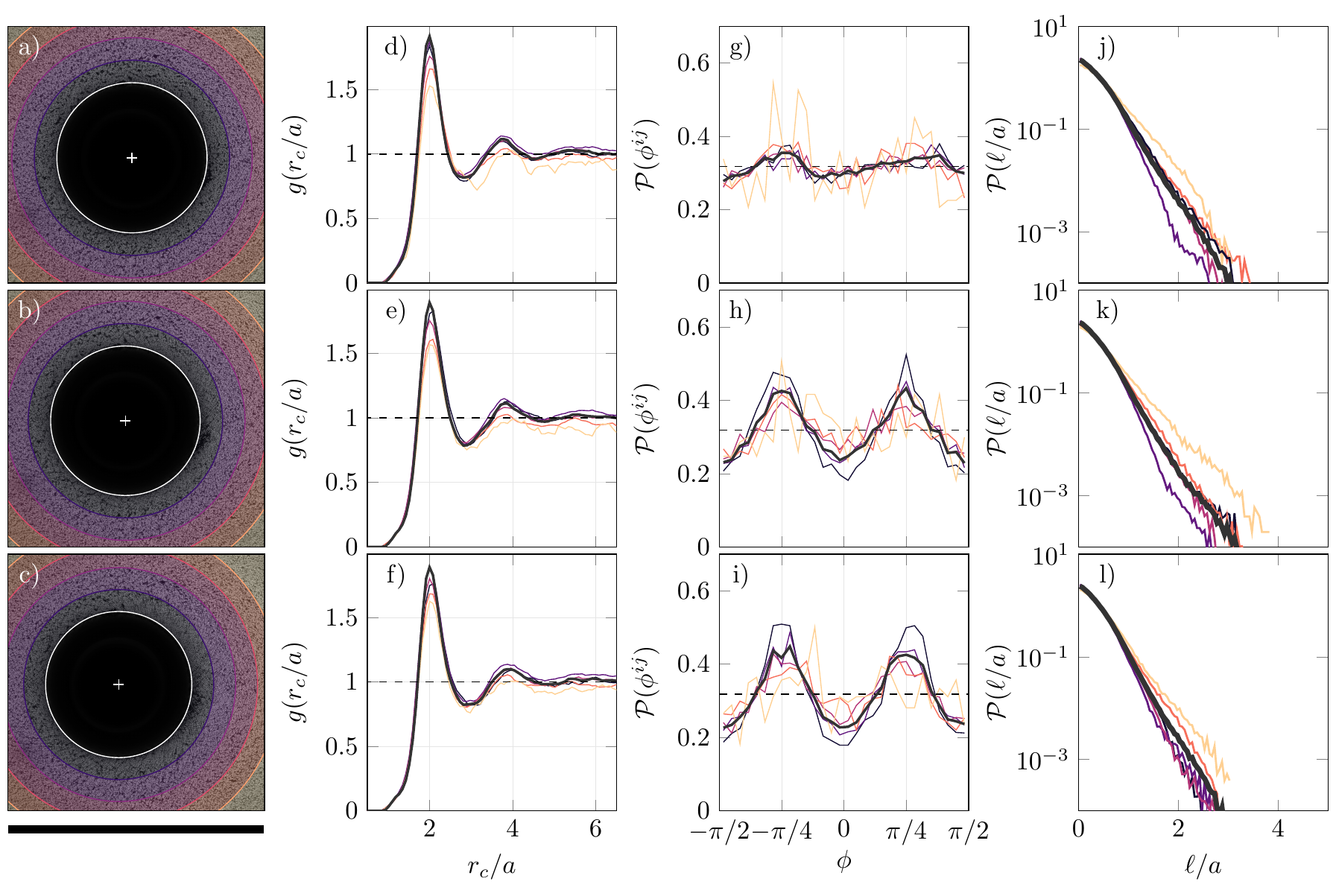} 
    \caption{Microstructural investigation of the gel around the bubble before and after an acoustic excitation. Panels (a-c) correspond to Panels (b-d) of Figure~\ref{fig:acoustics_observations} and overlays indicate sectors in $r$ over which structural quantities are averaged in other panels. The black scale bar is $300~\mu$m. Panels (d-f) show the global particle pair correlation function (thick, dark grey) and sector-averaged pair correlation functions (thinner, coloured lines). The inter-particle distance $r_c$ is normalised by $a$, half of the $r_c$ value where the first peak of $g(r_c)$ is observed before ultrasound is applied. Panels (g-i) correspond to the bond orientations $\phi^{ij}$ (see Figure~\ref{fig:A_DEFINITIONS} for more details), using the same colour conventions as in Panels (d-f). Panels (j-l) show the distribution of void sizes $\ell/a$ in the gel (see Figure~\ref{fig:A_DEFINITIONS} for details) using the same colour conventions as Panels (d-i).}
    \label{fig:acoustics_structure}
\end{figure*}

We point to the fact that the local particle ordering has no consequences on the local particle fraction around the bubble, at least within the accuracy of our confocal microscopy method (around 2\%), as shown in SI Figure~4(b). The observed ordering upon ultrasound-driven bubble dynamics is further quantified in Figure~\ref{fig:acoustics_structure}, which presents the evolution of the same structural indicators previously calculated for bubble dissolution. Panels (a-c) of Figure~\ref{fig:acoustics_structure} correspond to panels (b-d) of Figure~\ref{fig:acoustics_observations}. Comparing Figures~\ref{fig:acoustics_structure}(d-f), we do not notice any clear changes in $g(r_c/a)$ upon application of ultrasound.

The bond orientation parameter $\phi^{ij}$ shows a more dramatic change before and after the application of ultrasound: while the initial distribution is relatively flat and shows very little structuring, two clear peaks emerge at $-\pi/4$ and $+\pi/4$ after applying ultrasound. The peak locations seem to be preserved after the application of a second acoustic excitation and the peak amplitude increases slightly [Figure~\ref{fig:acoustics_structure}(i)]. Structuring is more pronounced for particles close to the bubble (purple lines) than for particles initially far from the bubble (pink-orange lines)  [Figure~\ref{fig:acoustics_structure}(h-i)] while the orientation effect is limited to a region $R \leq r \leq 2 R$. In contrast with the clear effect seen on the bond orientations, the void size distribution $\mathcal{P}(\ell/a)$ is not significantly altered before and after the application of the acoustic pulses [Figure~\ref{fig:acoustics_structure}(j-l)]. The main effect of ultrasound-driven bubble dynamics is therefore to reorient particles relative to the surface of the bubble.

\subsection{Microstreaming flow during ultrasound-driven bubble oscillations}
\label{sec:microstreaming}

\begin{figure*}[t]
    \centering
    \includegraphics{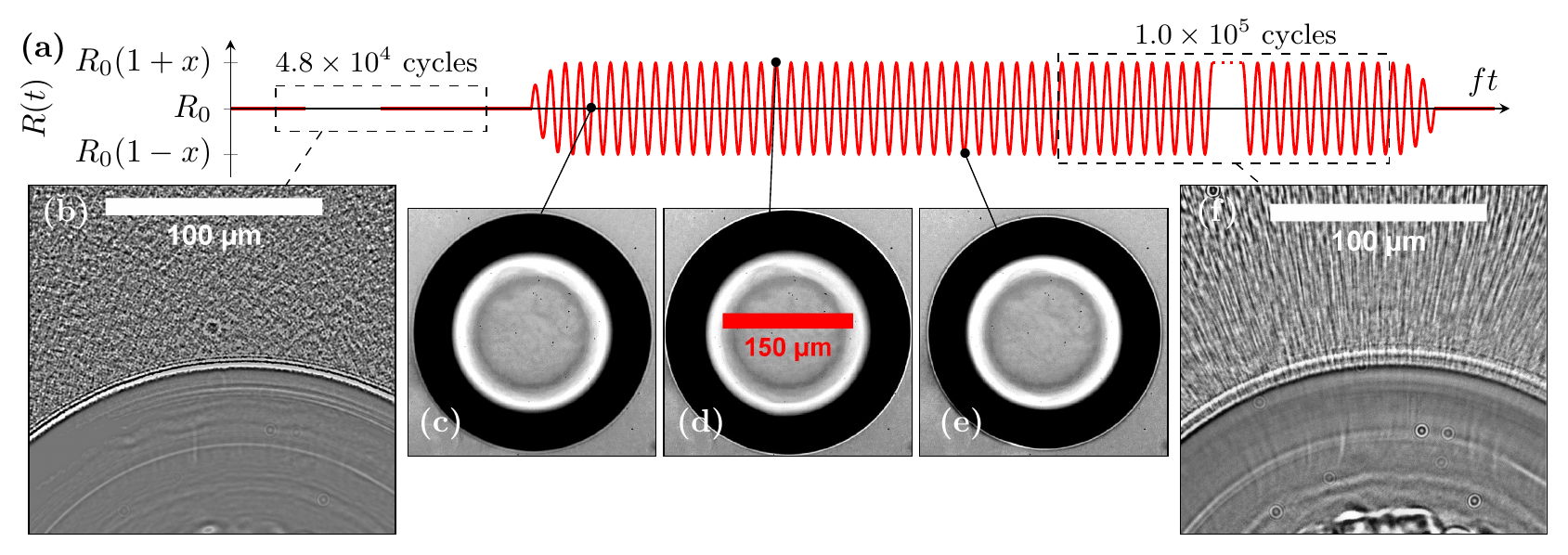}
    \caption{Microstreaming around an ultrasound-driven pancake bubble, see SI Video 2. (a) Schematic representation of the bubble time series. (b) A bubble and its surroundings at rest, imaged using the real-time brightfield setup and averaged over $N = 4.8 \times 10^4$ cycles. The bubble is located at the bottom of the picture. (c-e) Instantaneous pictures of bubble oscillations acquired with the high-speed brightfield imaging setup, close to its radius at rest [Panel (c)], at its maximal radius [Panel (d)] and at its minimum radius [Panel (d)] for a bubble oscillation amplitude $x = 0.03$. (f) Microstreaming observed when averaging brightfield images over $N = 1.0 \times 10^5$ cycles using the real-time setup .}
    \label{fig:microstreaming}
\end{figure*}

Brightfield imaging reveals a mechanism that is not visible in the low-frame-rate acquisitions with the confocal microscope and that can contribute to reorganise the gel particles during bubble oscillations: microstreaming, a secondary flow occurring on a much longer timescale than the period of bubble oscillations. Microstreaming around oscillating pancake bubbles confined between two planar walls has been reported \cite{MekkiBerrada2016} but has not yet been fully characterised or examined theoretically even in simple fluids. As shown in Figure~\ref{fig:microstreaming}, applying a continuous ultrasound excitation for around one minute ($N~\sim 10^{6}$ cycles) produces a microstreaming flow around the bubble. This is evidenced by the contrast between Figure~\ref{fig:microstreaming}(b) in which some particles of the gel may be resolved, whereas Figure~\ref{fig:microstreaming}(f) shows streak lines of aligned particles close to the bubble. SI video~2 highlights that microstreaming draws in randomly oriented clusters of particles above or below the equatorial plane of the bubble, which appear blurry as they are out of focus, and ejects the strings of well-aligned particles in focus, in agreement with the ordering noticed in the distribution of $\phi^{ij}$ in Figure~\ref{fig:acoustics_structure} (g-i) and the three-dimensional observations of Figure~\ref{fig:A_ZSTACK}. Close to the bubble, we measure a typical microstreaming velocity $v_\mu = 5~\mu$m\,.\,s$^{-1}$, yet this value is a lower limit since we cannot measure the $z$ (vertical) component of the velocity, which is expected to be large in close proximity to the bubble~\cite{Bolanos2017}. This velocity quickly decays one to two radii away from the bubble surface.

The high-speed time-aliased imaging (Section~\ref{sec:brightfield}) of SI Video 4 suggests that microstreaming is already active after $0.03$~s ($N \simeq 600$). We indeed expect a steady flow to develop over a distance $h$ in $N = f \tau = f h^2 \rho_{\rm l}/ \pi \eta \sim 5$ oscillation cycles from momentum diffusion, which is below the time resolution of SI Video 4 -- one frame every 200 cycles.

\subsection{Mechanisms governing gel restructuring}
During dissolution, bubbles apply to their surroundings a strain exceeding $\gamma_{\rm Y}$, which is necessary to locally break bonds in the gel, as seen from the particle rafts observed within the solvent pocket. Yet, bubbles apply this strain at a rate that is is extremely slow (of the order of $2.5 \times 10^{-3}$~s$^{-1}$), implying very small Péclet numbers: ${\rm Pe}^{\rm diss} \ll 1$ and ${\rm Pe}_{\rm dep}^{\rm diss} \ll 1$. In this case, the solvent may start flowing through the gel particle matrix\cite{Song2019}, breaking down the assumption of a homogeneous medium. We analyse this phenomenon in more detail in Appendix~\ref{appendix:bubble_dissolution} and we find that solvent flow through the particle matrix, while limited, can contribute to the solvent pocket formation. The slight increase in volume fraction seen at the solvent pocket edge [Figure~\ref{fig:dissolution_observations} and SI Figure 4(a)] is also compatible with scenario of  gel cluster densification and void size increase also reported in bulk oscillatory shear experiments~\cite{Moghimi2017} for low values of both ${\rm Pe}$ and ${\rm Pe}_{\rm dep}$.

During bubble oscillations, gel particles may rearrange both due to the primary fast oscillatory flow and the secondary, slower microstreaming flow. While both ${\rm Pe}^{\rm osc}$ and ${\rm Pe}_{\rm dep}^{\rm osc}$ are high, the maximum strain applied by the primary flow is given by the oscillation amplitude $x$, and no significant rearrangements are expected if $x$ remains below the yield strain~\cite{Moghimi2017}. The microstreaming Péclet depletion number ${\rm Pe}_{\rm dep}^\mu = v_\mu / (2 \pi R f x) {\rm Pe}_{\rm dep}^{\rm osc} = 0.3$ indicates that microstreaming close to the bubble surface is also a potential candidate to induce local gel ordering, in particular if $v_\mu$ is underestimated.

We hypothesise that microstreaming helps bringing particle aggregates close to the bubble surface to be internally restructured by the bubble oscillations, provided that their amplitude is comparable to or exceeds $\gamma_{\rm Y}$. Close to the bubble, the oscillations break down and reset the gel microstructure, as seen at high Péclet depletion numbers in bulk steady~\cite{Koumakis2015} and oscillatory shear flow in colloidal gels~\cite{Moghimi2017}.For our range of ${\rm Pe}_{\rm dep}^{\rm osc}$, the absence of change in average void sizes (or volumes) [Figure~\ref{fig:acoustics_structure} (j-l), i.e. the lack of increased gel structure heterogeneity is also compatible with results obtained in bulk, simple shear~\cite{Moghimi2017}. The microstreaming flow controls the extent of the rearranged region, around one bubble radius away from its surface. The gel structural properties are then modified and tuned at the local scale, near the bubble periphery. This scenario is however challenging to confirm, as we cannot measure the gel microstructure and the bubble oscillation amplitude simultaneously.

The particle alignment produced after gel rearranging ultimately remains puzzling, in particular the observed preferential orientation of the particles bonds $\phi^{ij} = \pm \pi/4$. This alignment is reminiscent of observations in a similar gel~\cite{Smith2007} under simple, oscillatory shear, where particles are found to align perpendicular to the flow, in the shear gradient direction. These directions are not so well-defined in extension, preventing us from drawing a more quantitative comparison with Ref.~\citenum{Smith2007}.

\section{Conclusions}
We have explored the potential for tuning the microstructure of a colloidal gel by exploiting different phenomena of bubble dynamics. By using a Hele-Shaw geometry, where the bubbles are confined into a ``pancake'' shape, we performed simultaneous observations of bubble dynamics and confocal microscopy of the microstructure of the surrounding gel. Bubble dissolution in such two-dimensional confinement applies predominantly a planar elongational deformation. We found that the rate of deformation during bubble dissolution is sufficiently slow that the solvent can partially flow through the soft-porous matrix of colloidal particles. This scenario leads to the formation of a solvent pocket around the dissolving bubble, which remains after complete bubble dissolution. The gel microstructure outside of the solvent pocket is, however, mostly unaffected by bubble dissolution. 

In contrast, high-frequency bubble oscillations driven by ultrasound at 15-21 kHz result in a local rearrangement of the microstructure with short-range order. This effect is visible after a few hundreds of cycles of ultrasound excitation, which corresponds to less than 0.1~s. We characterised the final microstructure by introducing an orientational order parameter that quantifies ordering around a circular domain. The average volume fraction of particles appears to remain unchanged. Using brightfield high-speed imaging we quantified the typical amplitude of ultrasound-driven bubble oscillations, and obtained direct observations of the dynamical phenomena in the surrounding gel. These observations reveal that multiple concurrent effects are at play during high-frequency bubble oscillations: in addition to the volumetric oscillations of the bubble, which apply an oscillatory planar elongation, a secondary flow, known as microstreaming flow and typical of bubble oscillations in confinement, was also observed. The resulting complex flow has components of both shear and elongation. The Péclet number is high for the oscillatory flow, which can then break particle clusters, and low for the microstreaming flow, which appears to only cause transport of clusters to the vicinity of the bubble. Neither flow has characteristics that directly explain the observed short-range order.  Particle-based simulations should help to dissect the roles of these different contributions and ultimately achieve control over the resulting microstructure. Future experiments should also address the evolution of the microstructure around unconfined bubbles. The results presented here pave the way to using ultrasound-driven bubble dynamics to tune the mechanical properties of colloidal gels at the local scale. Three-dimensional printing of deformable inclusions in colloidal gels could then be an promising route towards a locally programmable colloidal microstructure.

\section*{Acknowledgements}

The authors thank Esmaeel Moghimi and Mohan Das for the preparation and rheological characterisation of the colloidal gel, Andrew Schofield for providing the colloidal particles, and Pouyan Boukany for access to the confocal microscope. Discussions with Philippe Marmottant and Joost de Graaf are also acknowledged. This work is supported by European Research Council Starting Grant No. 639221 (V.G.) and EUSMI (Grant No. 731019).

\clearpage

\appendix

\section{Structural quantities}

\label{appendix:structure}

\begin{figure*}
    \centering
    \includegraphics[width=13.2cm, trim={0.5cm 1cm 0.5cm 1cm}, clip] {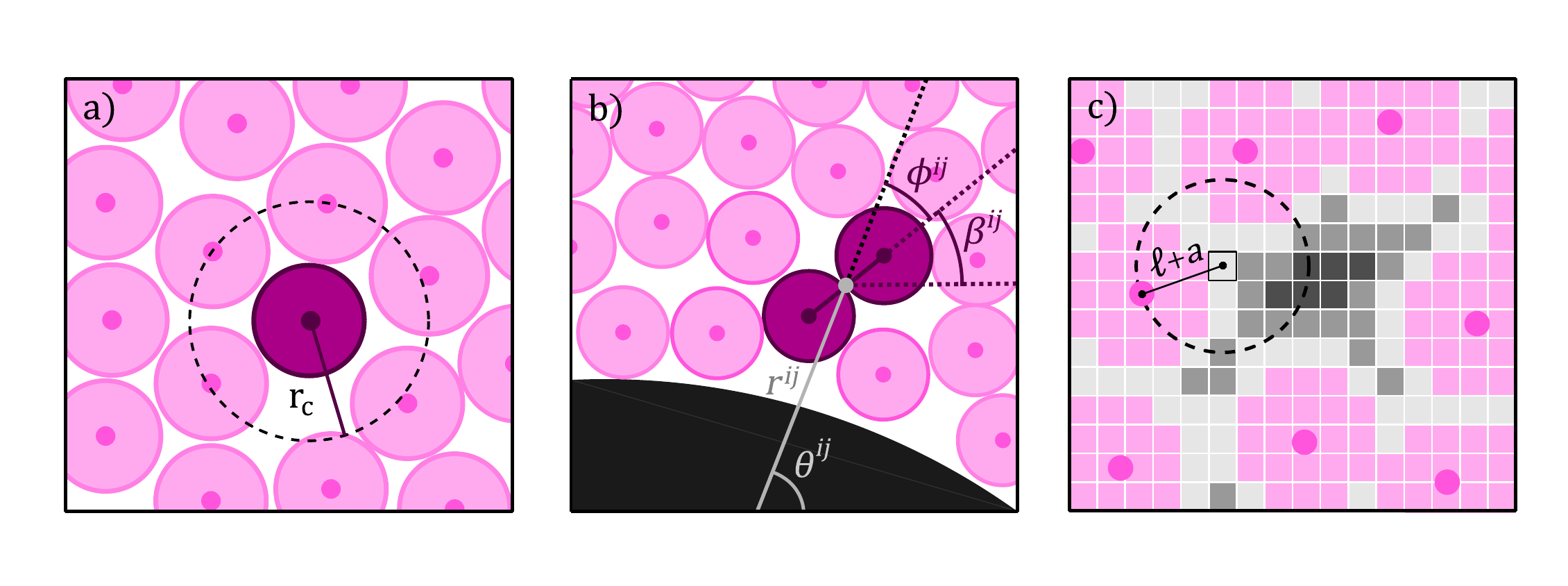}
    \caption{Structural quantities related to data processing. (a) Definition of the particle-particle distances $r_c$ used to defined the classical pair correlation function $g(r_c)$. (b) Definition of bubble-related quantities: distance $r^{ij}$ between a particle pair and the centre of the bubble, and angle $\theta^{ij}$ between the centre of the bubble and that of the particle pair, absolute bond orientation $\beta^{ij}$ and bond orientation $\phi^{ij}$. (c) Definition of the void sizes $\ell$ based on the estimated particle radius $a$ obtained computing $g(r_c)$.}
    \label{fig:A_DEFINITIONS}
\end{figure*}

Figure~\ref{fig:A_DEFINITIONS} shows the different structural quantities used throughout this article. Figure~\ref{fig:A_DEFINITIONS}(a) shows how we compute our pair correlation function, which is classically based on the distance $r_c$ from the centre of a given particle to all of its neighbours.

The bubble is a reference in the experimental setup from which several other structural parameters may be defined. We can draw the segment between the midpoint of neighbouring particles $i$ and $j$ and the centre of the bubble and define its length $r^{ij}$ and orientation $\theta^{ij}$, as seen in Figure~\ref{fig:A_DEFINITIONS}(b). Two particles are assumed to be neighbours when the inter-particle distance is below $1.08 a$, which then only include close neighbours. We define the absolute orientation of a particle bond through the angle $\beta^{ij}$. The difference between the absolute particle bond orientation $\beta^{ij}$ and $\theta^{ij}$ defines the relative bond orientation $\phi^{ij}$, simply referred to as the bond orientation in the main text, also shown in Figure~\ref{fig:A_DEFINITIONS}(b). In a purely radial particle arrangement around the centre of the bubble, we expect neither $\beta^{ij}$ nor $\theta^{ij}$ to present any structure, but we do expect $\phi^{ij}$ to show a strong maximum around zero. To avoid unnecessary duplicates in our definition of $\phi^{ij}$ (since $i$ is a neighbour of $j$ and conversely), we choose to constrain $\phi^{ij}$ between $-\pi/2$ and $+\pi/2$. 

In Figure~\ref{fig:A_DEFINITIONS}(c) we recall the definition of the void sizes $\ell$ in our confocal pictures, which matches that used in previous articles~\cite{Koumakis2015,Moghimi2017}. For each pixel of the acquired images, we compute its distance to the nearest particle. We then subtract $a$ from this distance and we name the result $\ell$. We ignore all pixels for which $\ell \leq 0$, which is equivalent to ignore all pixels inside detected particles. We see in Figure~\ref{fig:A_DEFINITIONS}(c) that the distribution of $\ell$ is very efficient at highlighting larger holes (regions devoid of particles) as they induce a significant broadening of this distribution.

\begin{figure*}
    \centering
    \includegraphics[width=17cm] {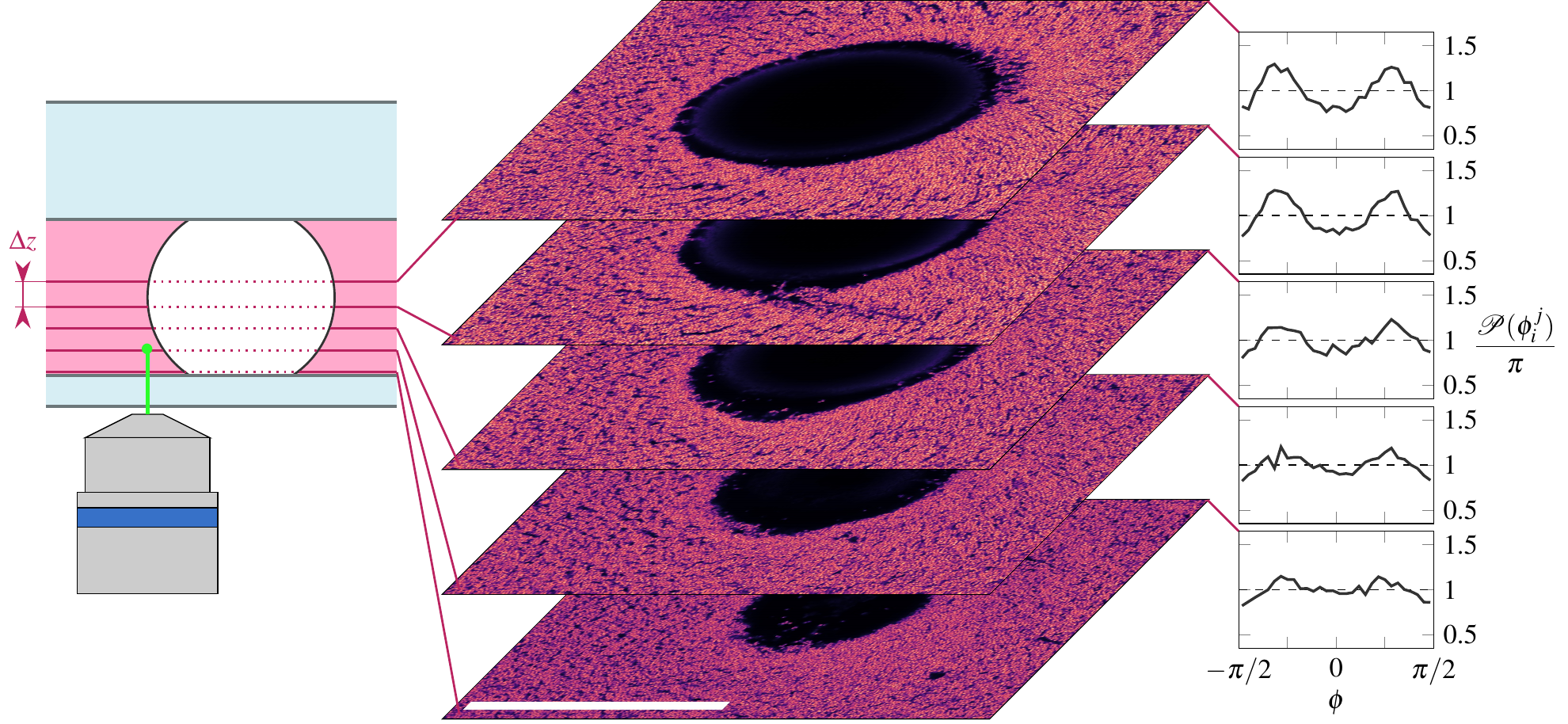}
    \caption{Z-Stack slices of the confocal gel particle after the application of repeated ultrasound excitation. The position of the slices on the left schematic accurately represent the depth at which they have been acquired in the sample, with all $\Delta z$ being identical and equal to $14.2~\mu$m. The corresponding bond orientation distributions $\mathcal{P}(\phi^{ij})/\pi$ are shown on the right. The scale bar is 100$~\mu$m}
    \label{fig:A_ZSTACK}
\end{figure*}

Figure~\ref{fig:A_ZSTACK} shows confocal images of the gel surrounding the bubble at different depths $z$ in the sample after ultrasound excitation. The normalised bond orientation distribution $\mathcal{P}(\phi^{ij})/\pi$ is plotted for each image in the $z-$stack. The microstructure close to the bottom wall is less ordered than at the equator of the bubble: the mechanism responsible for particle ordering (oscillations or microstreaming) is therefore maximal close to the bubble equator.

\section{Colloidal gel dynamics during bubble dissolution} 

\label{appendix:bubble_dissolution}

For the very small strain rates considered during bubble dissolution, the depletion Péclet  ${\rm Pe}_{\rm dep}^{\rm diss} \simeq 10^{-4}$ is exceedingly small, and viscous friction between solvent and gel cannot even bend the bonds between gel particles. Instead, the solvent may start flowing through the pores formed by the gel particle matrix, as observed in sparse fibrous networks~\cite{deCagny2016,Song2019}. A more precise dimensionless number $\mathcal{D}$ examines this phenomenon by comparing the linear elastic stress applied onto the material to the (Darcy) viscous drag applied by the solvent flowing through the matrix during bubble dissolution:
\begin{equation}
    \label{eq:darcy}
    \mathcal{D} = \frac{G}{\Delta p} = \frac{G a^2}{\eta \dot{R} R} f(\varphi)\,
\end{equation}
with $G = 10$~Pa the linear elastic modulus of the gel, $\eta= 0.012$~Pa\,s the solvent viscosity, $a= 0.66~\mu$m the particle radius, $\varphi = 44\%$ the particle volume fraction and $f(\varphi) = 2 (1-\varphi)^3/ 75 \varphi^2 = 0.024$ the dimensionless permeability of the gel. We assume here that the gel particles have a random, liquid-like arrangement~\cite{Carman1939} and ignore the --modest-- corrections associated with a fractal gel structure for $\varphi = 44\%$~\cite{Gelb2019}. Relative motion between the solvent and the particles becomes dominant when $\mathcal{D}$ falls below 1, or alternatively when $\dot{A} = |\dot{R}R|$, the absolute rate of change of bubble area, falls below $\dot{A}_{\rm crit} = Ga^2 f(\varphi) = 9 \times 10^{-12}$~m$^2$.s$^{-1}$.

\begin{figure*}
    \centering
    \includegraphics{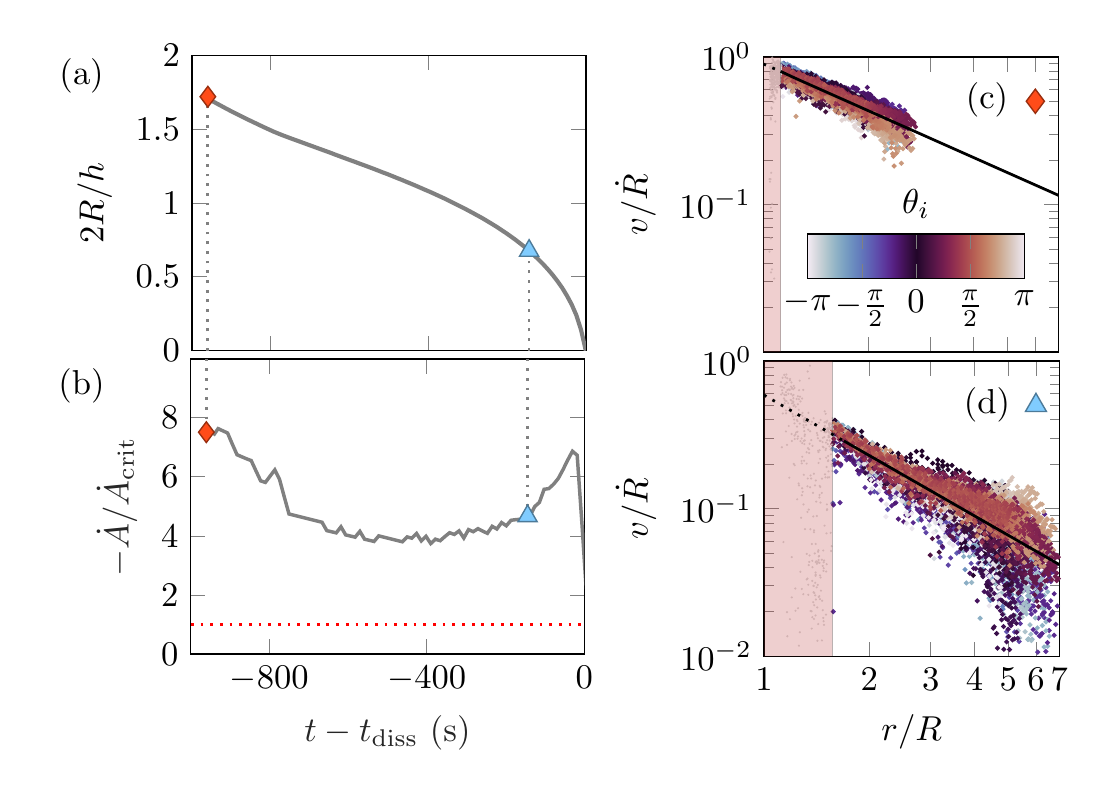}
    \caption{Bubble dissolution dynamics in the colloidal gel. (a) Normalised bubble size $2R / h$ as a function of the experimental dissolution time $t - t_{\rm diss}$ of a single bubble. (b) Bubble area loss $\dot{A}$ (normalised by $\dot{A}_{\rm crit}$) as a function of $t - t_{\rm diss}$. (c-d) Distribution of bubble velocity around the bubble for the frame pair marked by a diamond (c) and a triangle (d) of Panels (a) and (b). Experimental data in the light red region correspond to the solvent pocket devoid of particles and are ignored. The rest of the data points are fitted using a power law (solid black line, see Appendix text).}
    \label{fig:dissolution_dynamics}
\end{figure*}

Figure~\ref{fig:dissolution_dynamics}(b) shows $\dot{A}/\dot{A}_{\rm crit}$ for the bubble dissolution data of Figure~\ref{fig:dissolution_dynamics}(a). $\dot{A} / A_{\rm crit}$ first decreases from around $8$ to reach a local minimum of $4$ around $t-t_{\rm diss} \simeq 400~s$ and $R \simeq h/2$ before increasing again close to complete dissolution. Relative motion can be present, but will be small in our experiments: we expect it to be small close to the local minimum of $\dot{A}$ [triangle symbol in Figure~\ref{fig:dissolution_dynamics}(a-b,d)], and even smaller at the beginning of our experiment [diamond symbol in Figure~\ref{fig:dissolution_dynamics}(a-c)]. 

Figure~\ref{fig:dissolution_dynamics} (c-d) quantifies differential motion between the gel particles and solvent by computing the local particle velocity $v(r)$ using Particle Image Velocimetry routines~\cite{Liberzon2019}. We fit this field using a power law $v_r \left( r \right ) = - v_0 \left (r/R \right )^{k}$. Differential motion between particles and solvent is quantified by $v_0 / \dot{R}$, which is equal to 1 when particles and solvent move together, and less than 1 (in our case) otherwise. The fit quality is good and the velocity distribution is axisymmetric, as seen by the absence of dependence of $v_r$ with $\theta$ [coded with colours in Figure~\ref{fig:dissolution_dynamics}(c-d)]. Figure~\ref{fig:dissolution_dynamics}(c) shows a low, yet noticeable amount of differential motion ($v_0 \simeq 0.85 \dot{R}$) early during bubble dissolution and for larger $\dot{A}$, while Figure~\ref{fig:dissolution_dynamics}(d) shows higher differential motion ($v_0 = 0.60 \dot{R}$) later on and for smaller $\dot{A}$. This small differential motion is compatible with the solvent pocket formation and the slight particle compaction at edge of the solvent pocket during bubble dissolution, and more generally could also play a role in the cluster compaction and void size growth observed at low Péclet numbers in bulk shear~\cite{Koumakis2015,Moghimi2017}.

%% -------------------------------------------
%% -------------------------------------------
\bibliographystyle{tex/rsc}
\bibliography{bibliography}

\end{document}